\documentclass[aps,prb,reprint,groupedaddress]{revtex4-1}
\usepackage{graphicx}
\usepackage{dcolumn}
\usepackage{adjustbox}
\usepackage[dvipdfm]{hyperref}
\usepackage{hyperref}
\begin{document}
\title{Shear induced martensitic transformations in crystalline polyethylene: \\
direct MD simulation}
\author{I.A.~Strelnikov}
\author{E.A.~Zubova}
\email{zubova@chph.ras.ru}
\affiliation{%
N.N. Semenov Institute of Chemical Physics, Russian Academy of Sciences,
4 Kosygin Street, Moscow 119991, Russia
}
\date{\today}
\begin{abstract}
For the first time, we carry out molecular dynamics (MD) simulation of shear-induced martensitic phase transitions between the orthorhombic and non-orthorhombic (triclinic and monoclinic) phases of crystalline polyethylene (PE) in the framework of a realistic all atom model of the polymer. We show that the variation of the shear rate allows observing on a nano-sample both a strongly nonequilibrium phase transition occurring by random nucleation and irregular growth of a new phase ('civilian' way, for rapid deformations) and the coherent, or 'military', kinetics (generally considered as usual for martensitic transformations). We induce transitions from the orthorhombic to the triclinic phase according to two transformation modes observed in experiment on PE single crystals. Rapid deformation favors the transition directly to the triclinic phase, slow deformation - first to the intermediate monoclinic, and only then - to the triclinic phase. The second way corresponds to the experiment on extended chain PE. We explain this result and analyze the competition between different transformation and plastic deformation modes. Rotations of PE chains around their axes necessary for the transition between the orthorhombic and non-orthorhombic phases are executed by short twist defects diffusing along the chains. The transition between the monoclinic and triclinic phases occurs through half-chain-period translations of the chains along their axes, mostly collectively, as crystallographic slips.
\end{abstract}
\maketitle
\section{Introduction}

\begin{figure}
\begin{center}
\includegraphics[width=0.98\linewidth] {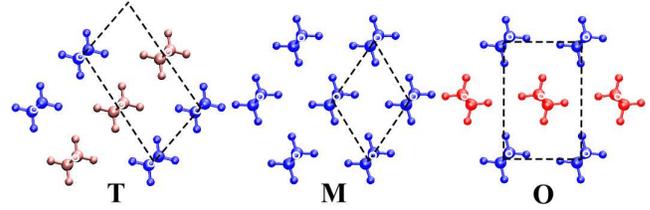}
\caption{Crystalline lattices known in alkanes and PE: triclinic (T, $A2/m$), monoclinic (M, $P2_1/m$), and orthorhombic (O, $Pnam$).
We show the projection onto the ($xy$) plane orthogonal to the chain axes. For the phases, we use the terminology by Kitaigorodskii \cite{1961-Kitaigorodskii} adopted in the book \cite{1973-Wunderlich-book-1} by Wunderlich. We use the term “triclinic” for the T lattice because its unit cell (containing two $CH_2$ groups of one chain) is triclinic, although a monoclinic crystallographic cell containing four $CH_2$ groups (belonging to two neighboring chains) is often used for the description of this lattice, and it is the projection of this cell which is shown in the figure.
\label{fig-PE-phases}
}
\end{center}
\end{figure}

Deformation induced martensitic transformations can be accompanied by twinning and generation of dislocations in the parent phase. This is the case in steel and other alloys, such as CoNi. Then the defects can prevent the reverse transition. In shape memory alloys the energy barrier for the phase transition is, on the contrary, much lower than for the plastic deformation modes. Under load, they are not actuated both in parent and product phases, and the transition is almost perfectly reversible. In some materials, the heights of these two barriers are comparable. In this case, the initiation of the transition can be impeded, and the behavior of the system under load can depend on many factors in a complicated manner. An example of such a transition is the transition from the orthorhombic (O) to the triclinic (T) phase in polyethylene (PE) (see Fig.~\ref{fig-PE-phases}).

When solidified from a melt under ordinary conditions, polyethylene (PE) forms a semicrystalline sample (about 70\% crystallinity), in which tiny O (high symmetry and high temperature phase) crystallites (lamellas) consisting of folded chains are embedded by 'tie' molecules into the amorphous matrix \cite{2006-textbook-Sperling}. The ground state of the system (an extended chain crystal) does not form because of kinetic reasons: long polymer chains have no time for it. In the lamellas, a load applied to the semicrystalline sample causes (in order of frequency) sliding on some crystallographic planes, twinning, and transitions of O crystallites into non-O phases.

The non-O phases are always present in commercial PE. A large amount of them was found in PE crystals formed on substrates \cite{1989-PE-epitaxial-crystallization,2004-PE-on-paper}, after polymerization inside nanochannels \cite{2005-PE-in-nanochannels} and in reactor powders \cite{2004-M-in-reactor-powders-Aulov,2018-M-in-reactor-powders}. These phases also appear under pressure, tension or shear, but the complete or almost complete transitions from the O to the non-O phases were observed only on specially prepared samples: on single crystals grown from a solution \cite{1964-Geil-single-crystals-stretching-direction,1974-Bevis-2of3-exp-single-crystals}, on extended chain (poly)crystals \cite{1987-alternative-Fontana,2007-Fontana-1of2}, or on (poly)crystals consisting of stacks of chain-folded lamellas \cite{1968-Seto,1974-Bowden-single-crystal-textured-PE}. In all these cases, the chains in all the crystallites were parallel.

High-symmetry O phase is preferable at high temperatures. If one deforms an O single crystal at 110$^\circ$C, the transition to the T phase does not occur \cite{1965-Geil-draw-temperature}. Annealing for 5 minutes at 100$^\circ$C of a semicrystalline sample containing T crystallites leads to their complete transition back to the O phase \cite{1968-Seto}. The single crystals that have transformed into the T phase, return to the O phase after removing the applied stress if they relax freely \cite{1965-mono-PE-single-crystal-stability}.

On the other hand, when a sample made of chain-folded lamellas is released after compression, only a part (which appeared at the late stage of deformation) of the formed T phase returns to the O phase, another part just twins \cite{1968-Seto}. In extended chain crystals, after a compression-decompression cycle even at temperatures more than 200$^0$C, the coexistence of the O and non-O phases is observed down to 2.3~GPa (under compression, non-O phases appear only at 6~GPa) \cite{2010-Fontana-2of2}. The very high temperature used by Fontana et al \cite{2010-Fontana-2of2} seems to be needed for the full recovery of the O phase under conditions of the experiment. From these facts, one has to conclude that there should be high barriers dividing the free energy minima corresponding to the O and non-O phases.

In addition, the picture is complicated by the existence of two non-O phases having higher free energies than the O phase: monoclinic (M) and T (Fig.~\ref{fig-PE-phases}). They both appear at compression of extended chain crystals at 280$^0C$ (M at 6~GPa, T at 15.5~GPa), they coexist at high pressures (40~GPa), and they both are present at decompression, the share of the T phase increasing with pressure decrease \cite{2007-Fontana-1of2,2010-Fontana-2of2}. It is possible that the M phase also appeared in the other experimental works cited above, but it was not detected because of similarity between the lattice parameters of the M and T phases.

\begin{figure}
\begin{center}
\includegraphics[width=0.90\linewidth] {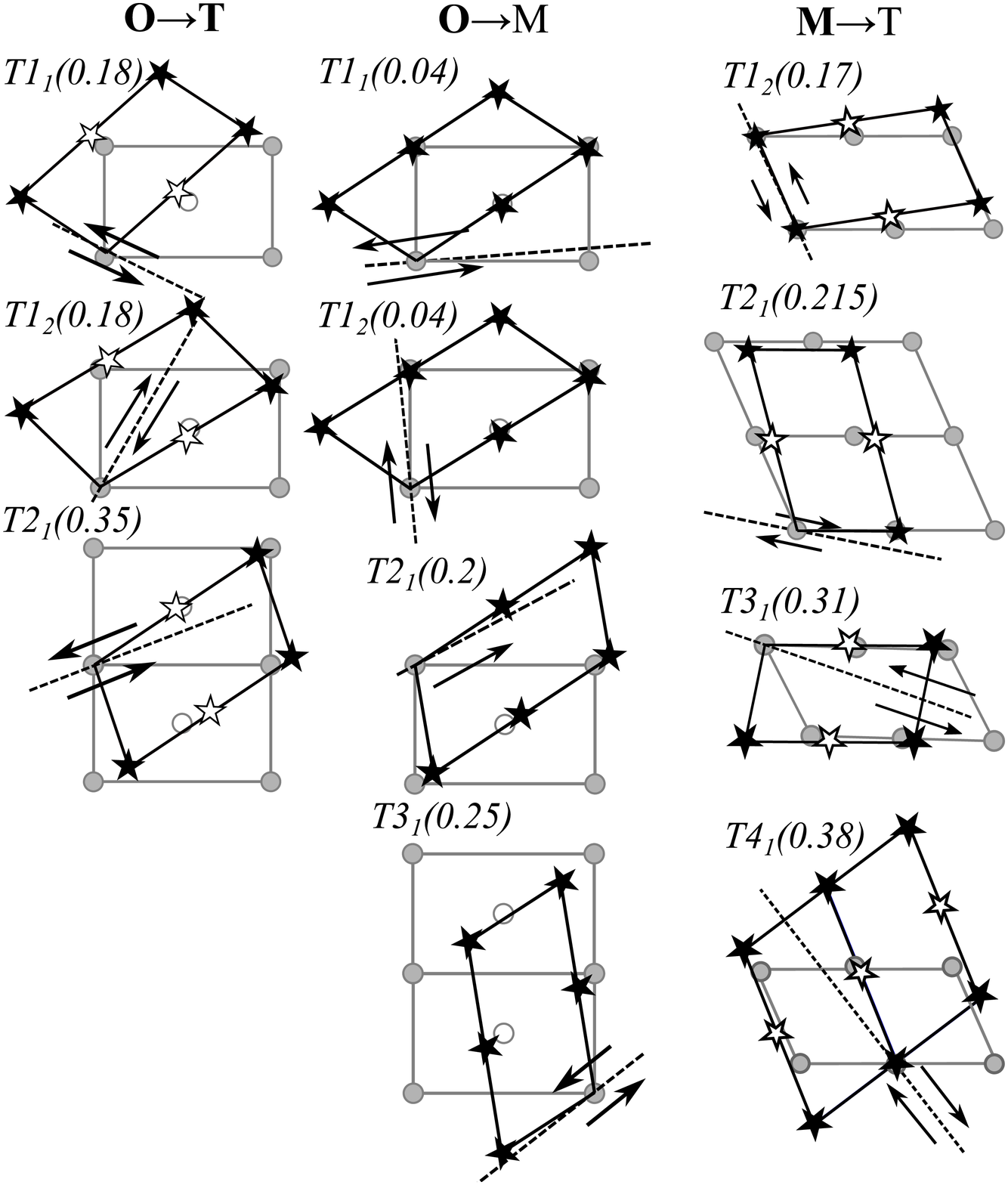}
\caption{Orientation relations of the parent and product phases in transitions between the M, T, and O phases. We show invariant planes (dashed lines) and shear directions calculated by method suggested by Acton et al \cite{1970-Bevis} for the lattices at 300K in the Amber force field. Bracketed quantities are the magnitudes of shear $s$ for the modes ($s$ equals approximately to the ratio of the shift of a molecule parallel to the invariant plane to the height above this plane). Here we showed only some particular modes, and (unlike Bevis et al \cite{1971-Bevis-geometry-of-phase-transitions-in-PE}) we numbered the modes dropping the unrealistic ones which do not shear all molecules to their correct positions.
\label{fig-transformation-modes}
}
\end{center}
\end{figure}

The densities of the O, T and M phases are very close in the wide range of pressures \cite{2007-Fontana-1of2}, therefore the transitions between these phases are considered to be martensitic. More exactly, the deformation needed for transitions between the phases is an invariant plane strain, almost a simple shear. Based on the observed diffraction patterns, Seto \cite{1968-Seto} determined the approximate relative orientation of the product T and parent O phases. The crystallographic analysis of the phases shows that there is a whole series of possible transformation modes bringing one lattice to another \cite{1971-Bevis-geometry-of-phase-transitions-in-PE}. The smallest displacements of molecules in $(xy)$ plane are provided by modes $T1_1$, $T1_2$ (relative locations of the molecules in these two modes are very close to each other and consistent with the conclusion of Seto's analysis \cite{1968-Seto}, but the invariant planes are different) and $T2_1$, $T2_2$ (see Fig.~\ref{fig-transformation-modes}).

In tensile deformations of monocrystals, modes $T1_1$ and $T2_1$ are operative
\cite{1973-Bevis-exp-monocrystals,1974-Bevis-exp-fold-sector-dependence}, while only mode $T1_2$ was observed
in compression experiment on single crystal textured (bulk) PE \cite{1974-Bowden-single-crystal-textured-PE}.
The further experimental investigation of the transitions between O and T phases in strained PE monocrystals was hindered by two unexpected problems. First, the mode of transformation proved to be dependent on fold sector (the pattern of returning of folded chains into the crystal) \cite{1974-Bevis-exp-fold-sector-dependence}, i.e. on boundary conditions. Secondly, as the martensitic transitions are caused by shear, the tension of a crystal induces too many competing modes of deformation, such as twinning and crystallographic slip \cite{1975-Bevis-multiple-deformation-processes}. The resulting picture was too complicated for productive theoretical analysis. Both the obstacles are also present for compression experiments on samples consisting of stacks of lamellae. They rotate and twin both in parent and product phases \cite{1968-Seto,1974-Bowden-single-crystal-textured-PE}. A successful shear experiment on any proper sample has never been reported. The reverse T$\rightarrow$O transition (induced by annealing) has been purposely studied only in one work \cite{1988-M-to-O-annealing}.

A molecular dynamics (MD) simulation of the transition can become a guide for the halted experimental studies. But,
although polymorphism is the most prominent feature of polymer crystals, its theoretical investigation is hindered
by the lack of adequate realistic models reproducing the phase transitions under study. Besides, the adequate statement of numerical
experiments is not evident as regards boundary conditions, shear directions, and sizes of samples.

Luckily, for PE, the simplest polymer, there is a phenomenological all atom force field, AMBER parm99 \cite{1999-AMBER-Parm99}, which reproduces all the three phases of PE, their energies being in the required order  (T is the ground state, O possesses the highest energy at 0K, the difference in energies and densities is very small, the barriers to transitions are rather high). All the phases are (meta)stable at normal conditions. Thus, it is possible to study the transitions between the phases in crystalline PE by direct molecular dynamics simulations, and to observe the kinetics of the transitions at the level of individual molecules.

In this work, we present a series of numerical experiments allowing to answer the most intriguing question on the O-T transition in PE: why is there the M phase between the parent O and the product T phases in the O-T transition \cite{2007-Fontana-1of2}? What prevents the direct transformation? Is there a way to actuate it?

\section{Details of MD simulations}

We wanted to simulate 'ideal' martensitic transitions between the lattices of crystalline PE. Namely, the transition should be induced by shear, not by tension or compression (as any PE lattice may be transformed to another by (almost) simple shear), and from the edges of samples. The samples should be small enough so that the stress field during the deformation at finite temperature was as homogeneous as possible, but the samples should be large enough so that the influence of the edges was minimal. We also wanted to avoid the dependence of the transition mode on the fold geometry of the polymer crystal \cite{1974-Bevis-2of3-exp-single-crystals}: chain folds can prevent some modes from being operative.

\begin{figure}
\begin{center}
\includegraphics[width=0.98\linewidth]{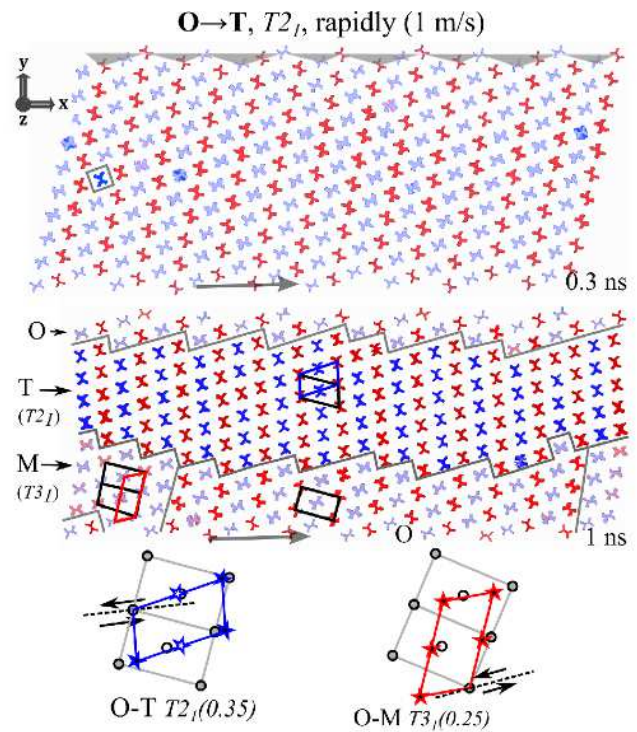}
\caption{(color online) Setting of MD experiments (simulation of O$\rightarrow$T transition according to mode O-T $T2_1$ in a small sample).
The plane orthogonal to $y$ axis is close to the invariant (shear) plane for mode $T2_1$.
The upper 'lid' (molecules on the gray substrate, which will not be shown on the subsequent figures) was fixed.
The bottom 'lid' (made of rigid molecules) was shifted in the direction of the  $x$ axis (the grey arrow).
The color of a molecule depends on its setting angle (the angle between the $x$ axis and the projection of the plane of the zigzag
...-C-C-C-... on ($xy$) plane). At the top of the figure: the first nucleus of the T phase (the blue chain within a square).
At the bottom: the final stage of the deformation. We show (in grey) the boundaries between the parent O and the product
 T and M phases. Comparing the relative orientations of the unit cells of the phases one can see
 (Fig.~\ref{fig-transformation-modes}) that the transformation modes are O-T $T2_1$ for the T phase and O-M $T3_1$ - for the M phase.
\label{fig-T21-small}
}
\end{center}
\end{figure}

Therefore, the samples consisted of finite chains (with their axes oriented along $z$ direction): 48 CH$_2$-groups (plus two terminal CH$_3$-groups) or 98 CH$_2$-groups (plus two terminal CH$_3$-groups). In one direction ($y$), the sample was restricted by two planes (orthogonal to the $y$ axis) approximately parallel to the invariant plane (shear plane K$_1$ \cite{1971-Bevis-geometry-of-phase-transitions-in-PE}) of the transformation mode under study. The shear direction in this plane $\eta_1$ coincided with the $x$ axis. To stabilize the samples under shear, we imposed periodic boundary conditions along the $x$ axis. One of the samples is shown in Fig.~\ref{fig-T21-small}.

The molecules belonging to the upper and the bottom edges of the sample formed two 'lids' by which the deformation was carried out. The molecules of both the lids were rigid zigzags with axes parallel to the $z$ axis. The molecules could shift along $z$ axis and rotate around it, but not around $x$ or $y$ axes. Such lids kept the sample intact.

In the process of shear, the axes of the molecules of one of the lids were shifted along the $x$ axis with a constant speed of 1, 1/4, or 1/16 m/s. They also could move along the $y$ axis independently of each other. The $x$ and $y$ coordinates of the mass centers of the molecules of the second lid were fixed, except for the action of a barostat which could change the $x$ coordinates of the mass centers of both the lids. Thus, the deformation we carried out was slightly different from a simple shear, but at this cost the sample had sufficient space to accomplish the phase transformations which requires, as we have mentioned in Introduction, not exactly the simple shear but an invariant plane strain.

To perform the MD simulations, we used the LAMMPS \cite{1995-LAMMPS} package \cite{LAMMPS-site} with the AMBER parm99 potential set \cite{1999-AMBER-Parm99} (see the force constants in Table~\ref{table-AA-potentials}). The time step was equal to 0.5~fs. The samples were kept at 300K with the use of the Langevin thermostat (the damping parameter of 1ps) applied to non-rigid chains. The Nose-Hoover barostat \cite{LAMMPS-NG-barostat,2004-NH-barostat} with the same damping parameter of 1ps kept zero stress along the x axis (S$_{xx}$ component of the stress tensor). The visualization was made with VMD \cite{VMD-site} support.
\begin{table}
\center
\caption{Potentials of interactions between the atoms (C - carbon, H - hydrogen) in the used all atom PE model
- the AMBER parm99 force field  \cite{1999-AMBER-Parm99}.}
\label{table-AA-potentials}
 \begin{tabular*}{0.48\textwidth}{p{0.1\textwidth}p{0.38\textwidth}}
 \hline
 \hline
\multicolumn{2}{l}{Valence bond potential: $\rm U(L)=K_L (L-L_0)^2$} \\
\hline
\multicolumn{2}{l}{C-C:             L$_0$=1.526\AA, K$_L$=310 $\rm kcal\:mol^{-1}$\AA$^{-2}$} \\
\multicolumn{2}{l}{C-H:  L$_0$=1.090\AA, K$_L$=340 $\rm kcal\:mol^{-1}$\AA$^{-2}$} \\
\hline
\hline
\multicolumn{2}{l}{Valence angle potential: $\rm U(\theta)=K_{\theta} (\theta-\theta_0)^2$} \\
\hline
\multicolumn{2}{l}{C-C-C: $\theta_0=109.5^{\rm o}$, $\rm K_{\theta}=40$ $\rm kcal\: mol^{-1} rad^{-2}$ }\\
\multicolumn{2}{l}{H-C-H: $\theta_0=109.5^{\rm o}$, $\rm K_{\theta}=35$ $\rm kcal\: mol^{-1} rad^{-2}$ }\\
\multicolumn{2}{l}{H-C-C: $\theta_0=109.5^{\rm o}$, $\rm K_{\theta}=50$ $\rm kcal\: mol^{-1} rad^{-2}$ }\\
\hline
\hline
\multicolumn{2}{l}{Torsion angle potential:} \\
\multicolumn{2}{l}{$\rm U(\varphi)=\sum_{n=1}^{3}K_{\varphi n} (1+cos(n\varphi+\varphi_{0,n}))$} \\
\hline
C-C-C-C: & $K_{\varphi 1}=0.2$ $\rm kcal/mol, \varphi_{0,1}=180^0$  \\
                 &$K_{\varphi 2}=0.25$ $\rm kcal/mol, \varphi_{0,2}=180^0$ \\
                 &$K_{\varphi 3}=0.18$ $\rm kcal/mol, \varphi_{0,3}=0^0$ \\
C-C-C-H:  &$K_{\varphi 1}=K_{\varphi 2}=0$  \\
                 &$K_{\varphi 3}=0.04$ $\rm kcal/mol, \varphi_{0,3}=0^0$ \\
 H-C-C-H: &$K_{\varphi 1}=K_{\varphi 2}=0$  \\
                 &$K_{\varphi 3}=0.0375$ $\rm kcal/mol, \varphi_{0,3}=0^0$ \\
\hline
\hline
\multicolumn{2}{p{0.45\textwidth}}{van der Waals pair interactions between atoms separated by more than 3 bonds or belonging to different molecules:} \\
\multicolumn{2}{l}{$U(r)=U_{LJ}(r)S(r)$} \\
\hline
\multicolumn{2}{l}{$U_{LJ}(r)=\varepsilon ( (R_{min}/r)^{12} -2(R_{min}/r)^{6})$} \\
\multicolumn{2}{l}{$\varepsilon_{CC}=0.1094$ $\rm kcal/mol$, $R_{min, CC}=3.816$\AA} \\
\multicolumn{2}{l}{$\varepsilon_{HH}=0.0157$ $\rm kcal/mol$, $R_{min, HH}=2.974$\AA} \\
\multicolumn{2}{l}{$\varepsilon_{CH}=(\varepsilon_{\rm CC}\cdot\varepsilon_{HH} )^{1/2}$;} \\
\multicolumn{2}{l}{$R_{min, CH}=(R_{min, CC}+R_{min, HH})/2$}\\
\hline
\multicolumn{2}{l}{ $S(r) = \left\{ \begin{array}{l}
1,\;r < {r_{in}}\\
\frac{{{{({r_{out}}^2 - {r^2})}^2}({r_{out}}^2 + 2{r^2} - 3{r_{in}}^2)}}{{{{({r_{out}}^2 - {r_{in}}^2)}^3}}},\;{r_{out}} < r < {r_{in}}\\
0,\;r > {r_{out}}
\end{array}\right. $ }\\
\multicolumn{2}{l}{$r_{in}= 12$\AA, $r_{out}= 13.5$\AA} \\
\hline
\multicolumn{2}{p{0.45\textwidth}}{For the atom pairs belonging to the same molecule and separated by 3 bonds, $S(r)$ was multiplied by 0.5}\\
\hline
\hline
\end{tabular*}
\end{table}
\section{Direct O to T transition}
For direct O$\rightarrow$T transitions, only three orientation relations were detected in experiment. They correspond to transformation modes \cite{1971-Bevis-geometry-of-phase-transitions-in-PE} $T1_1$, $T2_1 $, and $T1_2$ (see Fig.\ref{fig-transformation-modes}). We intended to implement modes $T1_1$ and $T2_1 $ known to be operative in single crystals of PE \cite{1973-Bevis-exp-monocrystals,1974-Bevis-exp-fold-sector-dependence}. The results of the MD simulations are summarized in table \ref{table-MD-experiments}.

We compared the transitions in large and small (in cross section) samples consisting of short and long chains. For the large samples, the rigid molecules in the lids constitute 10\% of the sample, and for the small samples - 20\%. One may expect that for the small samples the picture of the phase transition will be distorted (which turned out to be not the case), while in the large samples the stress field and the deformation will be inhomogeneous (which proved to be correct in some cases). We stopped the simulation when a plastic deformation of the formed composite samples started. The residual stress almost instantly relaxed upon release of the sample, without causing the reverse transition. We shifted the lid at four speeds - 1, 1/4, 1/8, and 1/16 m/s (the first speed corresponds to shear rate of 10$^8$s$^{-1}$ for the large samples). The transitions took place within 1-32 nanoseconds.
\begin{table*}
\caption{Results of MD simulations for the large samples, all but one consisting of (short) chains of 50 carbon atoms. The one of 100 carbon atoms is marked as 'long'. For all the velocities of the lid, we listed the observed modes, the resulting phase, the maximal (or minimal for negative stresses) value of S$_{xy}$  component of the stress tensor, the values of engineering shear strain at the beginning (g$_1$) and at the end (g$_2$) of the transition and the duration of the transition.
}
\label{table-MD-experiments}
\begin{tabular*}{1\linewidth}{c|c|c|c|c|c|c|c}
\hline
shear for:                            &vel.,      & observed modes   & result                 & S$_{xy}$,       &g$_1$, & g$_2$, & time,    \\
                                           &$m/s$   &in order of appearance &                 &MPa                  &$deg.$ & $deg.$ & $ns$\\[2pt]
\hline
 O$\rightarrow$T,    &          &                                                                                               &                             &        &&& \\
$T2_1$      &1      & O-T $T2_1$(0.35); O-M $T3_1$(0.25); T-M $T2_1$(0.215): all civilian      &  T+M(long)&185&-3.61 & -12.75&1.42\\
     &          &                                                                                                                                &   M             &162&-3.07&-8.24&0.744\\ [2pt]
     & 1/4    & O-M $T1_1$(0.04), civil.; O-T $T2_1$(0.35), military; M-T $T3_1$(0.31), milit.&T            & 150  &-2.23  &-16.48&9.18\\[2pt]
     & 1/16  & O-M $T1_1$(0.04), civil.; twinning of M, military; M-T $T1_1$(0.17), civil.           &  T           &193   &-4.29&-20.17&31.2\\[2pt]
\hline
O$\rightarrow$T,               &                       &  &                                               &      &         &&\\
 $T1_1$     &1       & O-M $T1_1$(0.04); O-T $T1_1$(0.18); M-T $T1_2$(0.17): all civilian & T    &-247 &2.85 &9.69&1.26 \\[2pt]
                 & 1/4    & O-M $T1_1$(0.04), civilian; M-T $T1_2$(0.17), military                      &  T    &-212 &2.45 &9.11&4.84\\[2pt]
\hline
T$\rightarrow$O, &                       &  &                                               &      &     &&    \\
$T2_1$               & 1/4   & O-M $T1_1$(0.04), civil.; T-O $T2_1$(0.35), military; T-M $T4_1$(0.38), milit.&  O+M &-63 &0&13.1&8.5 \\
with nucl.           &                         &  &                                               &        &          &&\\
\hline
\end{tabular*}
\end{table*}
\subsection{Transformation mode O-T {\it T2}}
 Mode $T1$ requires the smallest displacements of the molecules, while the transition according to mode O-T $T2$ should noticeably bend the crystal (see Fig.~\ref{fig-transformation-modes}), and, indeed, the vertical rows in the samples visibly curve during the shear. We start with mode T2, as it shows more variety in behavior depending on the deformation rate and the size of the samples.
\subsubsection{O-T {\it T2}, rapid deformation (1 m/s): random nucleation and irregular growth of the T phase, strong competitor: O-M {\it T3}}
\begin{figure*}
\begin{center}
\includegraphics[width=0.98\linewidth]{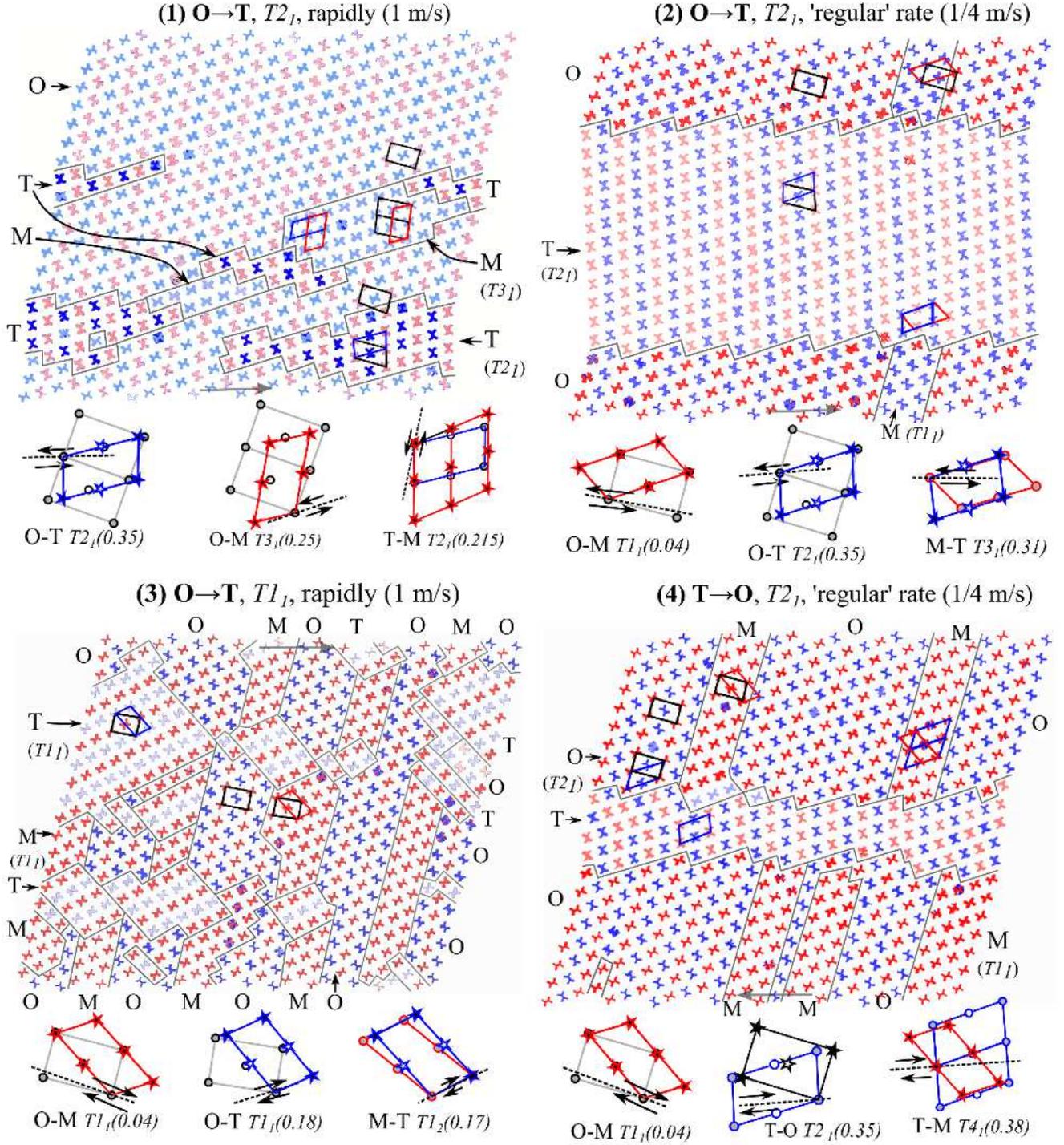}
\caption{Structure of the samples in the process of phase transitions. The observed transformation modes are shown below every sample. (1) Rapid O$\rightarrow$T $T2$ transition: competing T (mode $T2_1$) and M (mode $T3_1$) phases (sample with long chains). (2) 'Regular' O$\rightarrow$T $T2$ transition: 'military' propagation of the T phase; the domain boundaries are advancing to the upper and bottom edges of the sample, parallel to the invariant plane of the transformation mode $T2_1$. (3) 'Civilian' kinetics  of the rapid O$\rightarrow$T $T1$ transition: random nucleation and irregular growth of both the intermediate M and T phases. (4) The reverse T$\rightarrow$O $T2$ transition at 'regular' rate: parasitic M phase.
\label{fig-transitions}
}
\end{center}
\end{figure*}

Transformation begins with a turn of one chain through 90$^\circ$ (see Fig.~\ref{fig-T21-small}). The chain becomes a nucleus of the product T phase which grows with the deformation. As expected, the phase is oriented relative to the parent O phase according to transformation mode $T2_1$. During deformation, several nuclei appear and grow rather irregularly (in 'civilian' way \cite{1965-third-2002-Christian}): in different directions, but when they are small - preferentially along the $x$ axis, and after formation of a horizontal layer - in the $y$ direction. Rotations of the chains of the O phase happen more often near the T nuclei. Here arise M domains oriented according to mode O-M $T3_1$ (see Fig.~\ref{fig-transformation-modes}). In the small samples, it happens rather late, and the M phase does not noticeably spread. The result of deformation of the samples consisting of long chains and the samples consisting of short chains is more or less the same (Fig.~\ref{fig-T21-small}). Namely, the T phase (mode O-T $T2_1$) dominates. At the boundary of this phase, there are a domain of the M phase (O-M $T3_1$) and a remnant of the parent O phase. In the two small samples, the location of the non-T phases differed. It was near the upper lid for the sample with short chains (50 carbons), and near the bottom lid - for the sample with long chains (in both the cases, we shifted the bottom lid to the right).
In the large samples, the domains of the M phase (mode O-M $T3_1$) grow especially easily in the areas between the formed T nuclei (Fig.~\ref{fig-transitions}(1)).
%
In the large sample with long chains, the phases T and M ($T3_1$) spread out at similar rates, and, at the end of the deformation, share the sample approximately equally (see file O-T\_T2\_1\_(100CH2).mp4 in Supplemental Material \cite{movies}), and this ratio does not change if one continues the shear. In the sample with short chains the M phase ($T3_1$) emerges simultaneously with the T ($T2_1$) phase, propagates much sooner, and, in addition, easier consumes small T domains (by mode T-M $T2_1$). At the end of the deformation, the M phase ($T3_1$) occupies the whole sample (see file O-T\_T2\_1.mp4 in Supplemental Material \cite{movies}).
\subsubsection{O-T {\it T2}, 'regular' deformation (1/4 m/s): 'military' propagation of the T phase, weak competitor: O-M {\it T1}}
If we decrease the rate of the lid in the large sample with short chains to 1/4~m/s, the result of the deformation and the evolution of the sample are quite different (file O-T\_T2\_0.25.mp4 in Supplemental Material~\cite{movies}).
First, in the whole sample, only one nuclei of the T phase (mode $T2_1$) emerges and grows. Before its formation, a small domain of the M phase (mode $T1_1$)  appears. Later, it is being gradually occupied by the expanding T phase. Secondly, the growth of the T phase is not random, but very close to what can be described as a 'military' way \cite{1965-third-2002-Christian}. Formed in the center of the sample, the horizontal layer of the T phase expands to the lids, its boundaries being almost parallel to the invariant plane, and the sample deforming correspondingly (Fig.~\ref{fig-transitions}(2)). The result of the 'regular' (0.25~m/s) deformation is the expected pure T phase (mode $T2_1$), contrary to the rapid (1~m/s) shear of the same sample that has led to pure M phase (mode $T3_1$), which does not even appear at 'regular' rate.
\subsubsection{O-T {\it T2}, slow deformation (1/16 m/s): two intermediate M phases, twinning and crystallographic slips}
If the shear rate is further reduced by a factor of 4 (down to 1/16~m/s), the deformation results, as at velocity of  1/4~m/s, in a pure T phase (mode $T2_1$), but the transition flows differently  (file O-T\_T2\_0.0625.mp4 in Supplemental Material~\cite{movies}). The sample first transforms into the M phase oriented according to mode O-M $T1$. We saw one row of this phase at the velocity of  1/4~m/s. This phase grows not through a front propagation. The nuclei of the M phase arise mainly at the lids of the sample and represent one molecule turned through 90$^\circ$. The growth of the M phase occurs only along the vertical row in which the nucleus appears. Molecules flip one after another, starting from the lid. In the following deformation, the M phase changes its orientation to mode O-M $T2_1$ (relative to the parent O phase). The transformation goes not through gradual distortion of the cell geometry and gradual change of setting angles of the molecules. Instead, we saw the formation and development (in the 'military' way) of a twin of the first intermediate (O-M $T1$) M phase. The twin quickly caught all the sample. The following deformation of the second (O-M $T2_1$) M phase went through crystallographic slips along
directions in ($xy$) plane until, in the end, this phase transformed to the T phase according to mode M-T $T1_1$ (see Fig.~\ref{fig-transformation-modes}) by crystallographic slips parallel to the chain axis.
\subsection{Transformation mode O-T {\it T1}}
\subsubsection{O-T {\it T1}, rapid deformation (1 m/s): intermediate M phase, random nucleation and irregular growth of the T phase}
When the lids are parallel to the invariant plane of transformation mode O-T $T1_1$, the intermediate M phase (mode O-M $T1_1$) appears during the transition even at the high strain rate. The nucleation and growth of this M phase are similar to that of the intermediate M phase emerging in the slow deformation (1/16~m/s) according to mode O-T $T2_1$. After a noticeable spread of the domains of the M phase, it begins to transform to the T phase through the slip of half the chains along their axes. T domains emerge at random places and grow independently of each other, in 'civilian' way (see video file O-T\_T1\_1.mp4 in Supplemental Material \cite{movies}, till 0.89ns), and their lattices may not fit together. But when two 'incoherent' domains come into contact, one of them reorganizes (through the transition to the M phase) so that they form a whole (0.94ns). Thus, during the transition, the domains of all the three phases, O, M, and T, coexist, but it does not cause a noticeable deformation of the sample (Fig.~\ref{fig-transitions}(3)) because the lattice parameters for the properly chosen cells are very close at this orientation relation (according to mode $T1_1$). The result of the transition is an ideal T phase.
The kinetics of the transition in the samples composed of long and short chains is approximately the same. The only feature of the transition in the sample with short chains is that in it the nuclei of not M, but immediately T phase also occasionally emerge.
\subsubsection{O-T {\it T1}, 'regular' deformation (1/4 m/s): intermediate M phase, almost 'military' propagation of the T phase}
At a lower strain rate (1/4~m/s) in the large sample with short chains, the transition begins in the same way as at the high rate, with the growth of the M phase nuclei (according to the same O-M $T1_1$ mode) starting from the edges. However, the T phase begins to appear much later, when most of the sample is already in the M phase. The transition to the T phase starts mainly from one of the edges, the M phase and the remnants of the O phase being captured rather in 'military' than in 'civilian' way. Although the boundary of the T phase is noticeably different from a plane, but, as in the case of 'regular' deformation according to mode O-T $T2_1$, this boundary is marked by a slight but still visible curvature of the vertical rows in the sample. Thus, the slowing down of the shear leads, as for mode O-T $T2_1$, to more cooperative kinetics of the transition (file O-T\_T1\_0.25.mp4 in Supplemental Material \cite{movies}) and to the transition first to the intermediate (O-M $T1_1$) M phase, and only then to the T phase.
\section{Reverse T to O transition}
 To study the reverse T to O transition, we stopped the deformation in O-T $T2_1$ 'regular' (1/4~m/s) transition (the large sample with short chains) at the moment close to the finish and changed the velocity of the moving lid to the opposite (see file T-O\_T2\_0.25.mp4 in Supplemental Material \cite{movies}). As expected, the boundaries of the T domain went back to the lids. However, in the remainder of O phase, there appeared the common 'parasitic' M phase (O-M $T1_1$) that spread together with the O phase catching the T domain (there is the proper mode T-M $T_4$, see Fig.~\ref{fig-transitions}(4)). And we remember from the direct transitions that this direction of shear favors the transition of the O phase to this M phase on their boundaries parallel to the $y$ axis. Therefore, the sample after the shear was a mosaic of the O and M phases.
\section{Discussion.}
\subsection{Mode O-M T1: intermediate and parasitic.}
After the described simulations, we can answer the question posed at the end of the Introduction about the reason for the appearance of the M phase between the O and T phases in the compressed sample \cite{2007-Fontana-1of2}. Geometrical parameters of the M lattice are much closer to those of the O lattice than the geometrical parameters of the T lattice. Namely (see Fig.~\ref{fig-transformation-modes}), the M phase can be obtained from the O phase by a simple shear with a very small magnitude of 0.04 (mode O-M $T1$). An analogous (with the close orientation relation, $T1$) mode for the O-T transition has a magnitude of 0.18. Because the corresponding shear directions are fairly close, it is not unexpected that we have seen that the shear in the direction of mode O-T $T1$ causes a transition to the M phase according to mode O-M $T1$ at any shear rate.

Moreover, mode O-M $T1$ may, due to its tiny magnitude, lead to the growth of the M phase also under shear in any direction. Indeed, there are two modes of transformation O-M $T1$: $T1_1$ and $T1_2$, their invariant planes being almost perpendicular (see Fig.~\ref{fig-transformation-modes}). A simple shear in any direction along any invariant plane will have a positive projection on the shear direction of one of the modes $T1_1$ or $T1_2$. Therefore, under quasistatic shear in any direction in the O lattice, mode O-M $T1$ should be actuated. And the shear directions for two modes O-T with the smallest amplitudes $T1$(0.18) and $T2$(0.35) are very close to the shear direction for mode O-M $T1$.

Only in the rapid shear in the direction far from the direction of O-M $T1$, the M phase did not have time to be born before the deformation becomes large enough to actuate another mode with larger magnitude, but in a direction closer to the shear. In our simulations, we did not observe the mode O-M $T1$ only in the transition O-T $T2$ at a speed of 1~m/s (then the modes O-T~$T2_1$(0.35) and O-M~$T3_1$(0.25) appeared, see Fig.~\ref{fig-transitions}(1)). At both lower speeds 1/4 and 1/16~m/s this mode was present. It was even actuated as a 'parasitic' in the remnants (nuclei) of the O phase in the reverse T-O transition (Fig.~\ref{fig-transitions}(4)). This mode can be operative also in rapid shear, if the shear direction is closer to the needed direction: in the transition O-T $T1$ at the speed of 1~m/s, the mode O-M $T1$ was the first to appear.

 It is unlikely that in hydrostatic compression experiment by Fontana et al\cite{2007-Fontana-1of2}, local shear rates could be higher than in our simulations (see the estimates in section~\ref{section-rates}). Therefore, it is not surprising that at the beginning of the experiment the M phase was observed. It is equally natural that after the transition to the M phase, the crystallites under the action of the same stresses passed into the T phase: the direction of shear for the transition O-M $T1$ is close to the direction of M-T $T1$ (see Fig.~\ref{fig-transitions}(3)). Thus, it can be said that the path from the O to the T phase naturally lies through the M phase.
\subsection{Rate of shear and kinetics of transitions.}
\label{section-rates}
At first glance, the used velocities of the molecules of the lid of our nano-sample –1/16-1~m/s - and the corresponding shear rates – ($0.6\cdot10^7$-$10^8$)s$^{-1}$ - seem unrealistically large. Even under impact deformation, shear rate reaches only 10$^5$s$^{-1}$ at striker velocities of about tens of m/s. On the other hand, the speed of sound along PE chains (in our model) is 12, 2.6 or 1.9~km/s (depending on polarization), and in transverse directions – 2-4~km/s. The  velocities of the front propagation were about 0.55~m/s for mode $T2_1$ and 2.2~m/s for mode $T1_1$, which is three orders of magnitude lower than the speed of sound in this direction. The experimental estimate of the rate of growth of the martensite phase in iron under rapid heating is about 1~km/s (sound velocity is about 6~km/s) \cite{1978-Nishiyama-book-22}.

There are several competing deformation modes in the O lattice of crystalline PE. Shifts in crystallographic planes (especially parallel to $z$ axis), twinning, and the O-M $T1$ mode (0.04) have minimal magnitudes. The case of the transition O-T $T2$ (see Table~\ref{table-MD-experiments}) can illustrate the possible dependence of the transition kinetics and even of the product phase on the strain rate when several transformation modes can be actuated. One can assume that (if the boundary conditions allow it), under quasistatic shear, the first mode to be actuated will be the one for which the projection of this shear on its direction will first reach the needed magnitude.

At the lowest speed (1/16 m/s) of shear in the direction of mode O-T $T2$(0.35), we observed sequentially: a transition to the M phase according to the O-M $T1$ mode, twinning of the M phase, and a transition to the T phase according to M-T mode (mainly through shifts in crystallographic planes parallel to the $z$ axis). However, the end result was the expected T lattice with orientation relation corresponding to the mode O-T $T2$. When we increased the speed up to 1/4 m/s, the transition to the M phase according to the O-M $T1$ mode happened only in one vertical row, and, accordingly, the entire subsequent chain of transitions did not follow. Instead, the expected O-T $T2$ mode was actuated, and the transition went according to it (in military way).

With a further increase in speed up to 1~m/s, in addition to the mode O-T $T2$(0.35), the mode O-M $T3$(0.25) is also actuated, predominantly at the boundaries of the T domains. The T and M phases develop simultaneously. In this case (as opposed to all the other simulations), the result of the transition depends on the length of the chains in the sample. In the sample of short chains (50 carbon atoms), the mode T-M $T2$(0.215) is also operative, and the transition results in the M phase. In the sample of long chains (100 carbon atoms), the T and M phases share the sample. Thus, the result of this highly nonequilibrium transition is critically dependent on boundary conditions.

The shear rate (together with boundary conditions) also determines what kind of kinetics of the transition to expect: military or civil. For the transition O-T $T2$(0.35), a speed of 1/4 m/s was 'regular': it allowed to carry out this transition in standard military way without actuating modes with smaller magnitudes. The higher speed of 1 m/s led to the civil kinetics of the nucleation and growth of the T phase and the parasitic mode O-M $T3$. A lower speed of 1/16 m/s already gave time for the nucleation of the intermediate M phase according to the (low-amplitude) mode O-M $T1$.

For the transition O-T $T1$(0.18), which is closer to mode O-M $T1$ both in orientation relation and in magnitude, this intermediate M phase appears even at a speed of 1 m/s. The speed providing the military transition to the T phase has to be higher. The speed at which the direct nucleation of this T phase would go in civilian way is still higher. For this mode, we did not seek to find both the speeds because they are not currently available in experiment. The decrease in speed for the O-T $T1$ transition leads to the change in the kinetics of the second transition M-T $T1$(0.17) from the intermediate M to the final T phase: civilian kinetics at a speed of 1 m/s becomes almost military at a speed of 1/4 m/s.

The observed civilian kinetics is an example of a strongly non-equilibrium first order phase transition. The nanoscale domains of two or three phases coexist, separated by lengthy boundaries of complex shape (Fig.~\ref{fig-transitions}(1) and (3)). Formation of these boundaries requires more energy, which is provided by greater stresses (see Table.~\ref{table-MD-experiments}), accompanying the rapid transitions. The excess energy can be spent on formation of the boundaries between many nuclei of the new phase and the matrix.
\subsection{Boundaries between the phases: twist defects and tension defects.}
To accomplish a transition between the O and non-O phases, in addition to the displacements of molecules in the plane $(xy)$, their rotations around their axes are required. In transformation modes O-T and O-M $T1$ and $T2$, half of the chains in the sample have to turn through approximately 90$^0$. Both the transitions start with the generation of a T or M defect in the O lattice: a chain rotated counter-clockwise or clockwise through 90 degrees. The energies of these defects proved to be approximately equal to 0.2 kcal/mol per one CH$_2$-group. Both the defects correspond to local minima.

\begin{figure}
\begin{center}
\includegraphics[width=0.5\linewidth]{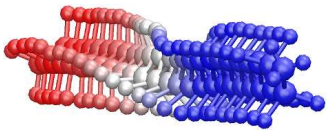}
\caption{
A twiston accomplishing localized rotation of a PE chain through 90 degrees.
\label{fig-twiston}
}
\end{center}
\end{figure}

\begin{figure*}[tb]
\begin{center}
\includegraphics[width=0.96\linewidth]{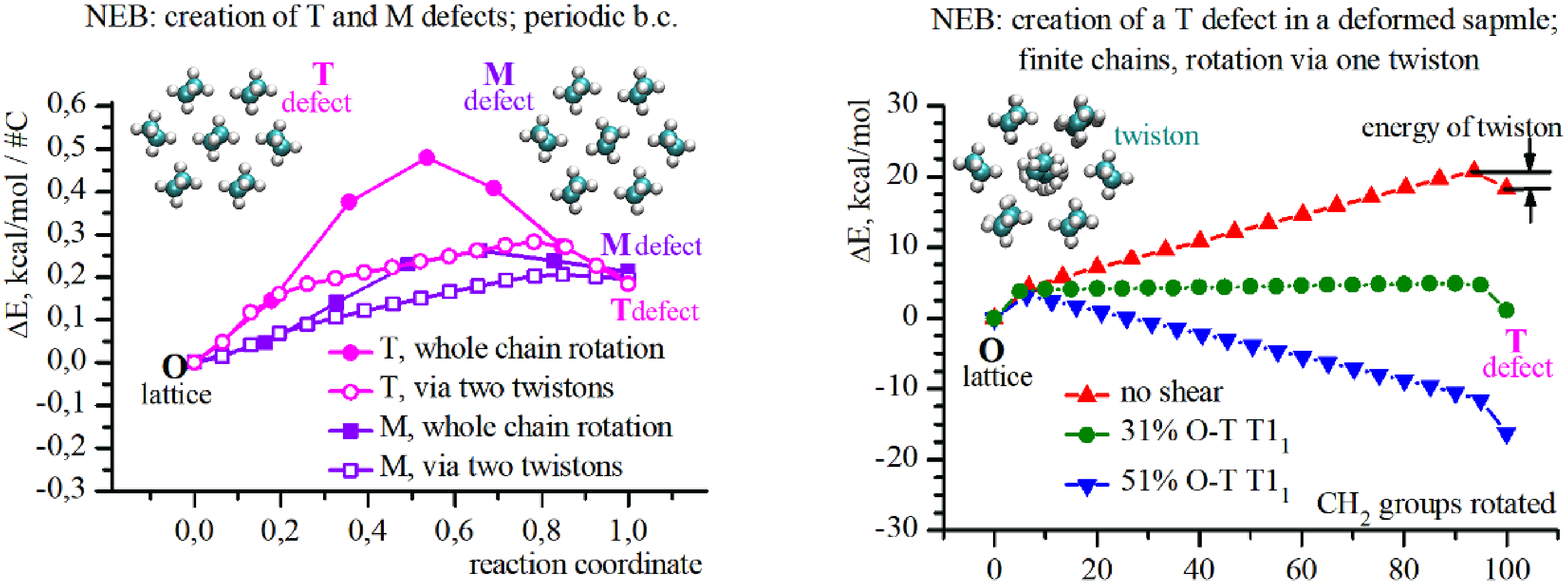}
\caption{
The energies of defects in O lattice with parameters corresponding to 300K. On the left: energies of T and M defects obtained under periodic boundary conditions with the chains consisting of 12 and 50 CH$_2$-groups (in the samples with 12 and 48 chains correspondingly). In the first case, the transition path lies through rotation of the chain as a whole. In the second case - through generation of a pair of twistons which move in opposite directions. After the twistons have been created, the energy of the system rises linearly as one CH$_2$-group rotates after another. When all the groups have been turned, the twistons annihilate. On the right: the dependence of the minimum energy path on shear.
\label{fig-energies-of-rotation}
}
\end{center}
\end{figure*}

Using the nudged elastic band (NEB) method \cite{1998-NEB}, we found that there is much lower barrier on the transition path to these minima through generation of a twiston (see Fig.~\ref{fig-twiston}) or, under periodic boundary conditions, a pair of twistons, than through  the rotation of the chain as a whole (see Fig.~\ref{fig-energies-of-rotation}, on the left).
Similar twist defects were observed in MD simulation of premelting in a PE crystal \cite{2012-PE-melting-Balabaev}.

The generation of one twiston turning a chain of O lattice to the position as in T defect costs 4~kcal/mol, and as in M defect - 2~kcal/mol. The barrier between the O lattice and the T defect is 20~kcal/mol for a finite chain with 100 carbons (Fig.~\ref{fig-energies-of-rotation}, on the right). But in a deformed sample, the energy of the T defect lowers with the shear. For example, under 31\% of the shear required for transition O-T T1$_1$, the energy of the defect is the same as the energy in the original position in O lattice. To turn a chain under this or greater shear, one needs to create only the twiston (see Fig.~\ref{fig-energies-of-rotation}, on the right). Therefore, in the described phase transitions, the upper boundary estimate for activation energy under shear equals to the energy of the twistons, 4~kcal/mol (for the direct O-T transition) or 2~kcal/mol (for the O-M transition).

The twistons emerge at the ends and sometimes (in pairs) in the middle of a chain. On long chains, we observed three twistons at the same time: a couple near the middle, and one near an end. The average overall width of a kink is 8-14 CH$_2$ groups (the core of the defect - about 6), and the average speed - 250-650~m/s (which allows to turn a chain containing 100 carbon atoms in 20-50 picoseconds).

Twistons are often generated at close time moments on neighboring chains. It should be noted that the speeds of the sound waves propagating along PE chains and polarized parallel and perpendicular to the axes of the chains are 12 and 2.6 (or 1.9)~km/s, respectively. We observed deceleration of the kinks and their movement in the opposite direction. In this case, the turn of a chain could not be completed, and the chain remained in the old phase. So, the motion of the twistons along the chains is a diffusion process, but it is a very quick diffusion.

Analogously to rotation around the axis, a long polymer chain in a PE crystal does not shift along the axis as a whole. A shift by a full chain period may be effected by motion of a vacancy (localized defect of tension) along the chain from one of its ends to another \cite{1999-vacancies-me}. The chain diffusion between amorphous and crystalline fractions in semicrystalline samples is implemented by motion of cheaper twist-tension defects (shift by half of a chain period and twist through 180$^0$ brings a CH$_2$ group into the position of the nearest neighbor) \cite{2007-chain-diffusion-me}.

In the process of transition between the intermediate M and the product T phases, we observed shifts of chains by half-chain-period. Namely, in the transition from the M to the T phase according to mode M-T $T1$(0.17) half of the chains in M lattice shifts by half-chain-period (which is equivalent to a turn through 180$^0$). These shifts are implemented by defects of tension. But the rigidity of a PE chain in tension (along its axis) is much more than the rigidity in torsion (around the axis). Therefore, the defects of tension are much wider than twist defects. For example, for the twist-tension defect responsible for the chain diffusion, the width of the area of tensile deformation is about 80 CH$_2$ groups, while the area of twist deformation - 25 CH$_2$ groups \cite{2007-chain-diffusion-me}. Therefore, when a chain is 50 CH$_2$ groups long, its shift looks as if the chain moves as a whole. The chain shifts very quickly, within a few picoseconds. We have often observed simultaneous displacements of several chains situated in one crystallographic plane. In this case, their cooperative shifts corresponded to crystallographic slips.
\section{Conclusion}
In the simplest polymer, PE, there are three crystalline phases (meta)stable under normal conditions, and able to transform one to another by shear-induced (first-order) phase transitions. There are many transformation modes ((almost) simple shears in different directions with different magnitudes) realizing these phase transitions. The height of the barriers for actuation of these modes is comparable to the height of the barriers for plastic deformation modes (crystallographic slips or twinning) \cite{1974-Bowden-single-crystal-textured-PE}. Thus, when a sample is deformed, the modes of deformation and of the phase transitions compete, and it is known that the result of the deformation depends not only on its type, but also on the structure of the sample. Another feature of phase transitions in polymer crystals is the micro-dimensions of single crystals or crystallites in semicrystalline samples and, accordingly, the strong influence of boundary conditions on these transitions.

An experimental study of the transition between the O and T phases of PE is hindered because of the need for special preparation of samples in which the competition with the deformation modes would be minimal. But even for such samples, the influence of the boundary conditions makes the productive theoretical analysis of the experiments very difficult. It is also not known whether there are another experimental conditions affecting the process of the transition. Many experimental results are unclear, in particular, the appearance of an intermediate M phase between the parent O and the product T phases in an extended chain sample under hydrostatic compression\cite{2007-Fontana-1of2}.

An MD study could serve as a guide for the halted experimental studies. Although for most polymers there are no force fields adequately modeling phase transitions between crystalline phases, the AMBER's force field \cite{1999-AMBER-Parm99} reproduces all crystalline phases of PE with correct relative energies.

In the present work, for the first time, in the framework of this realistic force field, we carried out an MD investigation of the phase transition between the O and T crystalline phases of polyethylene. We studied the first two (with minimal magnitudes) transformation modes which were experimentally observed. In our MD experiment, the transition was initiated by a shear along the invariant plane for the mode under study. The periodic boundary conditions were imposed only in one direction, and the chains had a finite length.

The study showed that the course and (to a lesser extent) the result of the transition depends on the rate of the shear deformation. A shear in the direction of mode O-T $T2$ (observed in single crystals) at high strain rate (velocity of the lid is 1~m/s) actuated not only the expected mode but also the transformation mode to the M phase O-M $T3$, and the transition went in civilian way: through random nucleation and irregular growth of the new phases. The result was (depending on the length of the chains in the sample) a combination of the T and M phases, or the M phase. At a lower speed (1/4~m/s), the same shear caused the coherent (military) transition to the expected T phase. The slow deformation (1/16~m/s) sequentially actuated the modes with smaller magnitudes and shear directions close to the shear direction of the deformation: O-M $T1$ (result: the M phase), twinning of this M phase (result: the M phase, oriented already according to mode O-M $T2$, similarly to the expected mode O-T $T2$), and finally, M-T $T1$ (result: the T phase, oriented according the expected mode O-T $T2$). So, the result of the slow deformation was the expected T phase oriented according to the mode we wanted to actuate, but the transition went through two differently oriented intermediate M phases.

The transformation mode to the first M phase O-M $T1$ has a very small magnitude of 0.04 (O-T $T1$ and $T2$ modes have magnitudes of 0.18 and 0.35, respectively). Because there are two nearly perpendicular directions of shear causing this mode, for any sufficiently slow simple shear, this mode will be actuated. For the shear in the direction of mode O-T $T2$, the critical speed was 1/4~m/s (one row of the M phase had already emerged, and was then absorbed by the T phase). Under the shear in the direction of mode O-T $T1$ (observed both in single crystals of PE and in bulk PE), the O-M $T1$ mode is being actuated even at a speed of 1~m/s. The resulting M phase transforms into the T phase by the same mode (M-T $T1$) as in the slow transition according to mode O-T $T2$. Thus, we see that the M phase (mode O-M $T1$) naturally emerges as an intermediate in the transition from the O to the T phase according to both modes O-T $T1$ and $T2$ observed in experiment.

The MD simulation also allows studying the transition process at the level of individual molecules, unavailable in experiment. In particular, we could observe the process of rotation of half of the chains through 90 degrees around their axes during the transition from the O phase to T or M. The turn proved to occur generally due to generation of a localized twist defect (twiston) on one end of a chain and the subsequent diffusion of the twiston to another end of the chain. The twiston is short (about 6 CH$_2$ groups), moves at speed of 250-650~m/s and rotates a chain containing 100 carbon atoms in 20-50 picoseconds. The energy of its generation is 4~kcal/mol for the O-T transition and 2~kcal/mol for the O-M transition. In this connection, one can notice that the close heights of the barriers for the studied phase transitions and crystallographic slips in polymers with chain diffusion between amorphous and crystalline fractions is not an accident. The reason is that the molecular mechanism of dislocation generation and motion also includes generation and motion of localized twist-tension defects \cite{2006-dislocations-in-PE,2017-Spieckermann-dislocations-in-iPP}.

In transition from the M to the T phase (or vice versa) according to mode M-T (or T-M) $T1$, the crystal needs not only a (small) deformation of its cell, but also a shift of half of its chains along their axes by half a period (or the rotation of the chains around their axes through 180 degrees). In our samples, consisting of finite chains 50 or 100 CH$_2$-groups long, we observed mainly collective half-chain-period translations corresponding to slip in a crystallographic plane. The shift of one chain occurs within a few picoseconds.

In the present work, only the first step is done in MD investigation of the phase transitions in PE. In particular, we did not study the reasons for the experimentally observed difficulty of the reverse transition to the O phase under normal conditions. We also did not analyze the initiation of the shear transformation modes, actuated along with the phase transitions in commonly used in experiment deformation processes: hydrostatic or uniaxial compression, or stretching. We also did not study the effect of boundary conditions (the length of chains and the direction of chain loops in single crystals and lamellae) on the actuation of a particular mode of phase transitions or deformations. We hope that our work will stimulate the research in these directions.
\begin{acknowledgments}
The work was supported by the Russian Science Foundation (award 16-13-10302).
The simulations were carried out in the Joint Supercomputer Center of Russian Academy of Sciences.
\end{acknowledgments}
\raggedright
\bibliography{AA_PE_def}
\end{document}